\documentclass[]{article}

\IfFileExists{latin1.sty}{\usepackage{latin1}}{\usepackage{isolatin1}}

\usepackage{graphicx}
\usepackage{fancyhdr}

\usepackage{url}
\usepackage{subfigure}
\usepackage{siunitx}
\usepackage{footnote}
\usepackage{url}
\usepackage{color}
\usepackage{amsmath}
\usepackage{amssymb}

\usepackage{tikz}
\usetikzlibrary{arrows,intersections}
\usetikzlibrary{shapes.geometric, arrows}
\usepackage{listings} %if lstlistings is used
\usepackage{changepage} %for changing topmargin on first page
\usepackage[figurename=Fig.,tablename=Tab.]{caption} [2008/04/01]
\author{Christoph Klemenjak\footnote{Alpen-Adria-Universit\"at Klagenfurt,
     Lakeside Park B10a, 9020
    Klagenfurt, christoph.klemenjak@aau.at} \, Peter
  Goldsborough\footnote{Technische Universit\"at M\"unchen, Arcisstrasse 21, 80333 M\"unchen, peter.goldsborough@tum.de}}

\title{Non-Intrusive Load Monitoring: A Review and Outlook}
\begin{document}
\maketitle
\setcounter{footnote}{2} %Auf Anzahl der AutorInnen setzen, damit die weitere Nummerierung der Fußnoten passt

\begin{abstract}
  With the roll-out of smart meters the importance of effective non-intrusive
  load monitoring (NILM) techniques has risen rapidly. NILM estimates the power
  consumption of individual devices given their aggregate consumption. In this
  way, the combined consumption must only be monitored at a single, central
  point in the household, providing various advantages such as reduced cost for
  metering equipment. In this paper\footnote{A shorter version of this paper was presented at the SKILL Students Conference 2016, part of the INFORMATIK2016 congress} we discuss the fundamental building-blocks
  of NILM, first giving a taxonomy of appliance models and device signatures and
  then explaining common supervised and unsupervised learning
  methods. Furthermore, we outline a fundamental algorithm that tackles the task
  of NILM. Subsequently, this paper reviews recent research that has brought
  novel insight to the field and more effective techniques. Finally, we
  formulate future challenges in the domain of NILM and smart meters.
\end{abstract}
%\begin{keywords}
%Non-Intrusive Load Monitoring, %Appliance Modelling, Smart Metering
%\end{keywords}

\section{Introduction} \label{sec:introduction}

Non-Intrusive Load Monitoring (NILM) techniques extract the power consumption of
single appliances out of aggregated power data. Given that a measurement device
employing NILM must only installed at a single point, none of the individual
appliances have to be equipped with metering devices. In Carinthia, Austria,
field trials using such technology are currently being implemented
\footnote{http://gewerbe.kelag.at/content/page\underline{ }strom\underline{
  }smartmeter.jsp}. These smart measurement devices, installed by energy
suppliers, pave the way for sophisticated disaggregation algorithms and possibly
also recommender-systems. Such a system would be able to detect devices that
have a need for maintenance and give appliance-specific feedback to the
consumer. Especially older household devices consume a lot more energy than new
ones. In \cite{benyoucef2010} measurements of an aged household refrigerator are
reported that consumed three times more energy than a new refrigerator. This is
a very specific example of a problem that could be solved by load-disaggregation
systems. Moreover, research has shown that appliance-specific feedback can save
up to 12 \% of annual power consumption \cite{CarrieArmel2013}. Data acquisition
units such as smart meters operate at a central point in the household's power
distribution network, generating measurement data of the total power
consumption. Currently available commercial smart meters were shown to have a
measurement deviation of 10-20 \% \cite{Zeifman2011}.This shows
clearly that actions have to be taken to improve future metering
units. \cite{klemenjak2014} states that the recent trend of metering units to
make use of more sophisticated energy monitoring ICs, providing more precise
measurements, has improved the effectiveness of NILM.

This paper is further structured as follows: Before discussing NILM algorithms,
the rudimentary concepts and fundamental vocabulary of the field will be
introduced. To investigate appliances, we must first classify them according to
their particular features and behaviour. This step is examined in Section
\ref{sec:appTypes}. Once measurement data is available, it must be analysed to
extract relevant appliance signatures, allowing tracking of said appliances. We
examine this topic in Section \ref{sec:signatures}. Naturally, a NILM system
requires to be aware about the circuits it is monitoring and thus to detect
them. For this, a wide range of learning approaches are applied, which we
discuss in Section \ref{sec:learning}. After touching upon these introductory
concepts, we describe a fundamental NILM algorithm published by Hart
\cite{hart:1992}. Since the publication of this algorithm in 1992, the field of
load-disaggregation has seen a tremendous amount of further research and novel
approaches. We present a selection of these ideas in Section
\ref{sec:newideas}. Lastly, in Section \ref{sec:conclusion}, we discuss future
challenges faced by the NILM community.

% Linie unter Kopfzeile To allow for comparison and numerical evaluation of an
% algorithm's performance, a set of metrics are needed, covered in Section
% \ref{sec:metrics}.

%\begin{figure}[h!]
%  \center
%  \includegraphics[scale=0.2]{feedback.png}
%  \caption{Possible annual energy savings \cite{CarrieArmel2013} }
%  \label{fig:appfeedback}
%\end{figure}

\section{Appliance types} \label{sec:appTypes}

Appliances differ in the number of operational states and their power
consumption behaviour. For the purpose of further discussion and analysis, we
therefore present three abstract models that are commonly used in research to
represent and characterise appliances. In detail, we will examine models for
on/off appliances, multi-state appliances and infinite-state appliances.

\subsection{On/Off appliances}

The first type of device is the so-called \emph{on/off} appliance. This class
includes common household appliances such as a toaster or a light bulb. Such
appliances consume only \emph{one} specific amount of power when active. For the
large part, on/off appliances are purely resistive. Appliances with a small
reactive part are assumed to be linear. A well-known fact is that electric power
is additive. This fact is exploited when describing a set of on/off
appliances. The total power consumed at time-instant $t$ is the sum of all power
signals $P_i(t)$. To modulate the power signal a switch process
$a_i(t) \in \{0,1\}$ is introduced. The product of the switch processes and the
power signals models the power consumption of a given appliance. The total power
$P_{total}(t)$ can therefore be estimated by:
\begin{equation}
  P_{total}(t) = \sum_{i=1}^{N} a_i(t) P_i(t) +e(t)
\end{equation}
The additive term $e(t)$ describes the deviation between the actual sum of the
modulated power signals and the measured total power. To estimate the state of
the appliances, the deviation $e(t)$ has to be minimised. In general, the
problem with this is that the complete set of power signals
${P_1(t) \dots P_N(t)}$ is not known. A second issue is that from a high
measurement uncertainty in estimating the total power $P_{total}(t)$, a bad
interpretation of the switching process may follow. Many appliances may be
estimated to be turning on and off at the same time. As a solution the
\emph{Switch Continuity Principle} was introduced in \cite{hart:1992}. It states
that in a small time interval the number of appliances changing their state is
also small. Consequently, we assume that in a small enough time window the
number of state transitions is zero. The sampling frequency of the acquisition
unit has thus to be high enough to detect such time windows. Between two such
intervals, in which the total power consumption is steady, appliances which
change their state can be identified.

\subsection{Multi-state appliances}

\begin{figure}[h!]
  \begin{center}
    \begin{tikzpicture}[scale=0.2]
      \tikzstyle{every node}+=[inner sep=0pt] \draw [black] (25.8,-16.9) circle
      (3); \draw (25.8,-16.9) node {$B$}; \draw [black] (28.8,-35.3) circle (3);
      \draw (28.8,-35.3) node {$A$}; \draw [black] (50.8,-27.7) circle (3);
      \draw (50.8,-27.7) node {$OFF$}; \draw [black] (26.852,-33.024) arc
      (-144.96699:-196.51257:15.552); \fill [black] (24.68,-19.68) --
      (23.97,-20.3) -- (24.93,-20.59); \draw (23.53,-26.82) node [left]
      {$+250W$}; \draw [black] (27.216,-19.543) arc (24.65565:-6.13521:24.428);
      \fill [black] (29.3,-32.34) -- (29.89,-31.6) -- (28.89,-31.5); \draw
      (29.83,-25.58) node [right] {$-250W$}; \draw [black] (28.547,-15.704) arc
      (108.56913:24.70218:17.312); \fill [black] (49.79,-24.88) --
      (49.91,-23.94) -- (49,-24.36); \draw (44.11,-15.68) node [above]
      {$-750W$}; \draw [black] (48.05,-26.51) -- (28.55,-18.09); \fill [black]
      (28.55,-18.09) -- (29.09,-18.87) -- (29.49,-17.95); \draw (41.71,-21.76)
      node [above] {$+750W$}; \draw [black] (49.807,-30.525) arc
      (-25.62228:-116.26232:13.75); \fill [black] (49.81,-30.52) --
      (49.01,-31.03) -- (49.91,-31.46); \draw (44.87,-38.18) node [below]
      {$-500W$}; \draw [black] (47.96,-28.68) -- (31.64,-34.32); \fill [black]
      (31.64,-34.32) -- (32.55,-34.53) -- (32.23,-33.59); \draw (36.62,-30.89)
      node [above] {$+500W$};
    \end{tikzpicture}
  \end{center}

  \caption{Finite state machine for an electric heater}
  \label{fig:heater}
\end{figure}
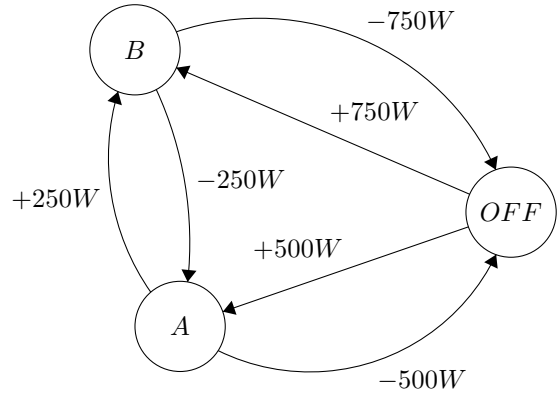

The second type of appliances are multi-state appliances, which have more than
one state of operation. Each of these states has a specific power consumption. A
common way to represent this class of devices is the finite state machine (FSM)
model. The graphic rendition of such a FSM consists of several circles, each
corresponding to a specific state of operation with a well-defined power
consumption. At the transition from one state to the other, visualised by an
edge, the power draw increases or decreases by the difference in consumption
between the two states of operation. As an example, let there be a finite state
machine model with two states of operation, as illustrated in Figure
\ref{fig:heater}. State A represents a power consumption of $\SI{500}{W}$ and
state B a consumption level of $\SI{750}{W}$. At the transition from state A to
state B the power consumption of the appliance rises with an amount of
$\SI{250}{W}$. In contrast to that the power consumption decreases by
$\SI{250}{W}$ from state B to state A. This is analogous to Kirchhoff’s law, as
the sum of the power changes is zero.

\subsection{Infinite-state appliances}

\begin{figure}[h!]
  \begin{center}
    \begin{tikzpicture}[scale=1]
    thick,
    >=stealth',
    dot/.style = {
      draw,
      fill = white,
      circle,
      inner sep = 0pt,
      minimum size = 8pt
    }
  ]
  \coordinate (O) at (0,0);
  \draw[->,thick] (-0.3,0) -- (10,0) coordinate[label = {below:Time}] (xmax);
  \draw[->,thick] (0,-0.3) -- (0,5) coordinate[label = {right:Power}] (ymax);
  \path[name path=x] (0.3,0.5) -- (6.7,4.7);
  \path[name path=y] plot[smooth] coordinates {(-0.3,2) (2,1.5) (4,2.8) (6,5)};

  \scope[name intersections = {of = x and y, name = i}]

\path       (1,3.75) -- node[below] { single } (3,3.75);
\path       (4,3.75) -- node[below] { multi } (6,3.75);
\path       (7,3.75) -- node[below] { infinite  } (9,3.75);

\draw [->,thick] (1,0) |- (1,2);
\draw [-,thick] (1,2) |- (3,2);
\draw [<-,thick] (3,0) |- (3,2);

\draw [-,thick] (7,0) -- (7,2.5);
\draw [-,thick]  (7,2.5)  parabola [bend pos=0.5] (9,1);
\draw [-,thick] (9,1) -- (9,0);

\draw [->,thick] (4,0) -- (4,2);
\draw [-,thick] (4,2) -- (5,2);
\draw [->,thick] (5,2) -- (5,3);
\draw [-,thick] (5,3) -- (6,3);
\draw [->,thick] (6,3) -- (6,0);

  \endscope

%   \foreach \y/\ytext in {2/500,3/750}
 %       \draw (1pt,\y cm) -- (-3pt,\y cm) node[anchor=east] {$\ytext$};

\end{tikzpicture}
  \end{center}

  \caption{Power consumption of different appliance types}
  \label{fig:shSchem}
\end{figure}
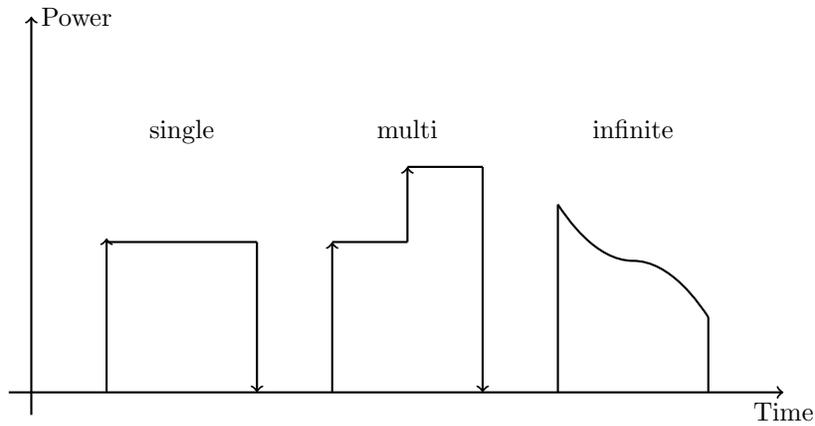

Lastly, there exist also appliances whose observable set of states is not
finite. For example, the power consumption of light-dimmers changes continuously
with no consistent step change. Such infinite-state appliances represent a
challenge to model and identify. Figure \ref{fig:shSchem} shows the power
consumption of such a continuously-varying power consumption. 
While on/off
appliances as well as multi-state appliances change their power consumption in
one clear and observable step, infinite-state appliance's power draw shows a
smooth pattern. One subclass of infinite-state appliances are those continuously
consuming energy, even when set in standby. Examples of such devices include
fire detectors or TVs.

%\begin{figure}[h!]
%  \centering
%  \includegraphics[width=1\columnwidth]{ppa.pdf}
%  \caption{Power consumption of different appliance types \cite{Zoha2012}}
%  \label{fig:shSchem}
%\end{figure}

\section{Appliance signatures} \label{sec:signatures}

Appliance \emph{signatures} describe characteristics specific to certain
devices, which can thus be used to identify and classify them. The importance of
such signatures was already pointed out in \cite{ting2005}, where a taxonomy of
voltage-current signatures was introduced to classify appliances. In general,
appliance signatures can be seen as measurable parameters, which provide
device-specific information extracted from physical quantities. Another taxonomy
was introduced in \cite{hart:1992}, where two classes of non-intrusive appliance
signatures are described: \emph{steady-state signatures} and
\emph{transient-state signatures}.

\subsection{Steady-state signatures}
\begin{figure}[h!]
  \centering
  \begin{tikzpicture}[only marks, y=.5cm]
    \draw plot[mark=*,xshift=0cm, mark size=0.8] file {scatter.data};
    \draw[->,thick] (0,0) -- coordinate (x axis mid) (8,0);
    \draw[->,thick] (0,0) -- coordinate (y axis mid)(0,12);

    \foreach \x/\xtext in {1/400,2/800,3/1600,4/2000,5/2400,6/2800,7/3200}
      \draw [xshift=0cm](\x cm,1pt) -- (\x cm,-3pt)
    node[anchor=north] {$\xtext$};
      \foreach \y/\ytext in {1/250,2/500,3/750,4/1000,5/1250}
      \draw (1pt,\y cm) -- (-3pt,\y cm) node[anchor=east] {$\ytext$};
      \node[below=0.5cm] at (x axis mid) {Active Power in W};
      \node[left=1.25cm,rotate=90] at (0,9) {Reactive Power in VAr};

      \draw (0.5,0.5) node[ellipse, minimum height=0.5cm,minimum width=1.2cm,draw] {};
      \draw (3,6) node[ellipse, minimum height=0.5cm,minimum width=0.5cm,draw] {};
      \draw (4,2) node[ellipse, minimum height=0.5cm,minimum width=0.7cm,draw] {};
      \draw (2,10) node[ellipse, minimum height=0.5cm,minimum width=0.5cm,draw] {};
      \draw (6,5) node[ellipse, minimum height=0.7cm,minimum width=1cm,draw] {};

      \draw (0.6,8) node[ellipse, minimum height=1.4cm,minimum width=1cm,draw] {};
    \end{tikzpicture}
    \caption{Distribution of appliances in a traditional P-Q plane}
    \label{fig:PQplane}
\end{figure}
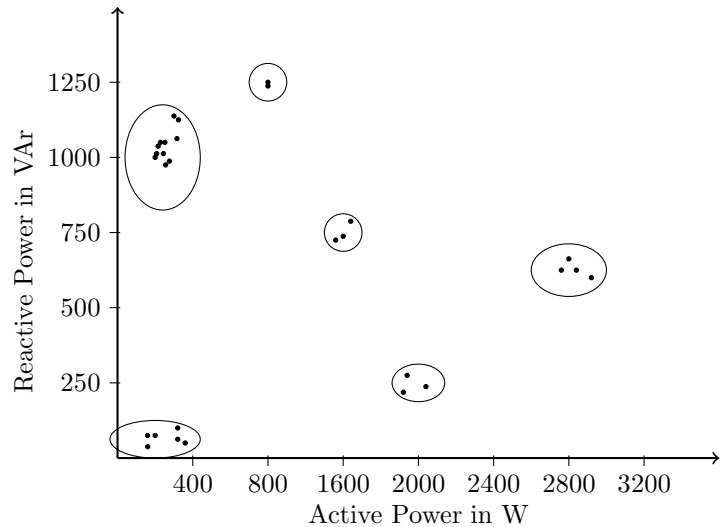

Steady-state signatures comprise features extracted from appliances when they
are not currently transitioning between two states but are operating at a steady
level of power consumption. More specifically, a steady-state signature is the
result of analysing the difference in certain characteristics between two steady
states of operation. Such a characteristic may be, but is not limited to, the
change in power consumption as was depicted in Figure \ref{fig:heater}. In
general such features can be categorised into the following groups:
\begin{itemize}
\item \textbf{Power Change: }Real and reactive power are the physical quantities
  of greatest interest, since they provide very characteristic information about
  appliances. To detect such features, the power consumption is estimated and
  plotted as shown in Figure \ref{fig:PQplane}. One major difficulty associated
  with this is the fact that certain power signatures may \emph{overlap}. This
  overlap results in a bad detection probability especially for appliances with
  low power consumption. Implementations such as \cite{huang2011classification}
  implement rely on these signatures.
\item \textbf{V-I Features: }The problem with overlaps can be solved by adding
  additional information about the appliances. By analysing the V-I
  characteristics, for instance the root-mean-squared (RMS) values of voltage
  and current, appliances with a similar power consumption may be further
  described and distinguished.
\item \textbf{V-I Trajectory: }Another method, using current and voltage signals, is to
  classify devices by extracting features out of the V-I trajectory. The shape
  of this trajectory shows useful characteristic features such as asymmetry,
  looping direction, and enclosed area. A recent application of analysing theses
  features can be found in \cite{Hassan2014}.
\item \textbf{Harmonics: }In \cite{hart:1992}, Hart states that analysing the harmonics of
  a device's current waveform by means of a Fourier Analysis can provide
  additional information about an appliance's characteristics. In particular, it
  was found that some non-linear appliances such as motors or light-dimmers
  produce current waveforms containing a specific set of harmonics, which can
  further aid in classification.
\end{itemize}

Extraction of such steady-state signatures does not necessarily demand for
high-end metering hardware. RMS values of current and voltage as well as frequent
power readings provide a good basis to extract steady-state signatures. Already low-cost
hardware such as introduced in \cite{klemenjak2014} can be used to identify steady-state
signatures.

\subsection{Transient-state signatures}

Situations exist where two different appliances may have very similar power
consumption profiles, reducing chances of correct identification of either
device. In such cases, examining the transients, i.e. the consumption behaviour
of an appliance when \emph{transitioning} between two steady states of
operation, can provide vital information. The transient signature of an
appliance is strongly influenced by the physical task it performs
\cite{Laughman2003}. For instance, the turn-on current of a computer system
differs massively from a lighting system due to charging capacitors. The shape,
size, and duration of such a transient can thus aid in distinguishing between
two appliances when their steady-state signatures alone do not provide a
sufficient basis for identification. At the same time, as noted in
\cite{hart:1992}, it must be considered that such momentary transition events
are less easily detectable than steady-state operation and may require a higher
sampling frequency. Another approach takes emitted voltage noise into
account. Each appliance in state operation transmits noise back to the power
line. This noise can be measured and categorised into on/off transient noise,
steady-state line voltage noise and steady-state continuous noise
\cite{Zoha2012}.

%\begin{figure}[h!]
%  \center
%  \includegraphics[scale=1.2]{transientpower.eps}
%  \caption{Transient behaviour in power change \cite{najmeddine2008state}}
%  \label{fig:transient}
%\end{figure}

\subsection{Ambient Appliance Features}

Both steady-state and transient characteristics are extracted from what are
arguably the most obvious sources: the power, voltage or current draw of an
appliance. However, in recent research, appliance-specific patterns have also
been extracted from environmental, ambient or behavioural sources. Such
techniques exploit the external impact of appliances, such as their
heat-dissipation or light-emission. For this purpose,
\cite{berges2010leveraging} advocates the collection of data from environmental
sensors. An implementation of this idea was proposed in
\cite{berg2011appliance}, where the deployment of electromagnetic field
detectors (EMF) to combine information about energy wastage and
power-consumption profiles was examined. In the same spirit,
\cite{guvensan2013energy} discusses the fact that home appliances emit sound
waves (noise). A system is suggested that correlates information about energy
consumption with sound recordings of the respective appliance. In contrast to
information provided by sensors about appliances directly, there exists also the
paradigm of Context-Aware Power Management (CAPM) \cite{dey2001conceptual}. CAPM
techniques typically examine signatures not necessarily extracted from
appliances themselves, but from their environment, users or usage behaviour. For
instance, \cite{kolterjaakkola2012} explores behavioural patterns including
duration of use and time of day. In \cite{kolterjaakkola2012} and
\cite{CarrieArmel2013} it is stated that such contextual information may also
include location or even weather patterns. Furthermore, \cite{priyantha2008tiny}
studies appliance-user interaction to facilitate load-disaggregation. The
behaviour and presence of human beings is traced by a set of motion sensors in
the building and combined with other NILM techniques. To gather such ambient
data, a wireless sensor network was proposed by \cite{dey2001conceptual}.

\subsection{Optimal Sampling Frequency for Signature Detection}

The number and kinds of steady-state and transient appliance features
recognisable from an aggregated consumption sample is strongly connected to the
frequency of measurement in the earliest stages of disaggregation. As noted in
\cite{kolter2011}, a wide range of utilised sampling frequencies is reported in
past literature. Many of the steady-state appliance features described above,
such as current harmonics or V-I trajectories and especially transient features
are more realistically attainable at higher sampling frequencies. With
\emph{high} frequencies we usually mean several kHz, although select approaches
have even employed MHz readings \cite{Gupta:2010}. High-frequency sampling rates
not only allow more fine-grained and detailed analysis of device-signatures, but
are also more flexible. The obvious benefit of having more samples available
than too few, is that when high-resolution data is not required or too bulky to
store, it can always be down-sampled to lower frequencies. On the flip side,
metering hardware for high sampling rates is practically non-existent in
households today, making NILM techniques requiring sampling rates in the region
of $\SI{1}{Hz}$ more practical and immediately applicable. $\SI{1}{Hz}$ readings
as used by Hart's algorithm \cite{hart:1992} allow for reasonably effective
examination of active and reactive power measurement. More recently, attempts
have been made to better adjust NILM algorithms to the low-frequency sampling of
conventional smart meters. For example, \cite{kolter:2010} makes attempts to
perform load disaggregation using discriminative sparse coding techniques on
power samples provided only on an hourly basis.

\section{Learning Approaches} \label{sec:learning}

Learning approaches for NILM can fundamentally be divided into \emph{supervised}
and \emph{unsupervised techniques}. The distinction between a supervised and an
unsupervised algorithm is whether or not ground-truth data about individual
appliance features is available to train the algorithm. If such device-specific
information is present, meaning that the algorithm knows a priori about the
appliances it is monitoring, the learning approach is limited to disaggregation
only. On the other hand, an unsupervised algorithm need not only perform
load-disaggregation, but additionally detect which appliances exist in the
circuit it is monitoring.

\subsection{Supervised Learning Approaches}

Supervised approaches feed the system with existing device-specific information,
such as its power consumption profile. This data may either already exist, such
as in the case of the REDD dataset \cite{kolter2011}, or is the result of an
initial training phase, in which a database of appliances and their signatures
is collected \cite{Aiad201696}. The actual load-disaggregation is commonly
performed by one of two techniques: \emph{Optimisation} or \emph{pattern
  recognition}. We will elaborate on either approach in the following
paragraphs.

\begin{itemize}
\item \textbf{Optimisation}: A straightforward method to solve the load
  disaggregation task is to model it as optimisation problem. Obtaining the
  solution for such problems is well-researched and builds on a simple
  concept. The extracted appliance features are compared to an existing database
  consisting of appliance features. When the deviation between the database's
  entry and the extracted feature can be minimised, the best match is obtained
  \cite{Aiad201696}. For a small number of appliances, this approach may very
  well be feasible. However, as discussed in \cite{Egarter2015} , the performance
  of this method deteriorates with an increasing number of loads, while the
  complexity increases. Another weak point of this approach is that it may have
  significant difficulties in distinguishing between loads with overlapping
  signatures.

\item \textbf{Pattern Recognition}: This approach detects appliances by means of
  clustering and mapping state-changes to a feature space \cite{Zoha2012}. An
  example of such clustering is given in Figure \ref{fig:PQplane}. As outlined
  by Hart in \cite{hart:1992}, the identified appliance features in the PQ plane
  are divided into clusters. Given this initial separation, the clusters are
  compared to those already known to the supervised system. In further detail,
  \cite{Zoha2012} identifies two main approaches: Bayesian classifiers and
  heuristic methods. For the former, it is assumed that two operating states of
  an appliance are independent of each other. While research has shown promising
  results for the Bayesian approach, the independence of states is clearly an
  ideal but not practical model. For example, the power state of a computer
  monitor usually depends directly on the power state of the connected computer.
\end{itemize}

\subsection{Unsupervised Learning Approaches}

Supervised learning approaches require an initial training phase and input of
external, labeled data. Practically speaking, for the average household, such
data does not exist. Therefore, unsupervised learning approaches, which are able
to operate without a priori information, are a promising
alternative. Unsupervised disaggregation techniques are required not only to
perform load-disaggregation, but must further train themselves
\emph{online}. This means that appliances need to be identified and extracted
from the aggregate power signal and their models added to the database of
existing devices. The quality of the load-disaggregation is thus additionally
dependent on the ability of the system to correctly identify existing
devices. Methods of probabilistic analysis such as Hidden Markov Models (HMMs)
and extensions thereof are especially suited to this task \cite{Aiad201696}. An
HMM is a probabilistic graphical model that differs from standard Markov models
in that the states are not directly observable, but can only be estimated
probabilistically given certain observations. For NILM purposes, an appliance
can be described as an HMM with $n$ hidden states $S = \{s_1,...,s_n\}$
representing the appliance's states of operation. Also, we define an observation
or emission matrix describing the probability for the appliance to be in a
certain state $s$ at time slice $t$ given the observation (emission) of an
aggregate power consumption signal. Lastly, there exists a transition matrix
$T = (a_{i,j}) \in \mathbb{R}^{n \times n}$ where $a_{i,j}$ represents the
likelihood for a transition of the appliance from state $s_i$ to state $s_j$
between two time slices $t$ and $t + 1$. More specifically,
$a_{i,j} = P(x_{t+1} = s_j | x_t = s_i)$ with $a_{i,j} > 0$ and
$\sum_{j=0}^n a_{i,j} = 1$. Factorial Hidden Markov Models (FHMM) are an
extension of the basic HMM. An FHMM models not only a single but many
independent hidden state chains in parallel, with the emission (the aggregate
power consumption) being thus a function of all states combined. In
\cite{Egarter2015} it is stated that this can help reduce the number of
parameters maintained by the system.

\section{Hart's NILM algorithm}

The algorithm introduced by Hart in \cite{hart:1992} is considered fundamental
in the NILM community and is the basis of many of today's load-disaggregation
techniques. For this reason, we will outline its basic operation briefly. The
general concept of Hart's algorithm is to meter a household's aggregate power
consumption, identify appliances and then track their behaviour. The algorithm executes the following tasks:
\begin{enumerate}
\item \textbf{Measure Power and Voltage:} Measurements of the aggregate power
  and RMS voltage signal are recorded at a sampling frequency of \SI{1}{\hertz}.

\item \textbf{Calculate Normalised Power:} The estimated power signals are
  normalised (smoothed) depending on the power line voltage. This allows for
  immediate comparison of power levels.

\item \textbf{Edge Detection:} An edge-detection algorithm is applied to the
  normalized power signals. This algorithm extracts steps in power consumption
  and labels the time instants.

\item \textbf{Cluster Analysis:} The output of the edge detection algorithm is
  used to create points in the PQ-plane. Points nearby are clustered.

\item \textbf{Build Appliance Models:} From the clusters obtained finite state
  machine (FSM) models are created. The simplest state machine is an on/off
  appliance consisting of two symmetrical events at the PQ-plane.

\item \textbf{Track Behaviour:} The estimated appliance models are
  tracked. Whenever a modelled appliance performs a state transition, the
  algorithm recognises this behaviour.

\item \textbf{Tabulate Statistics:} Statistics and characteristics of the models
  obtained so far are calculated and tabulated. These statistics may also be
  used to predict the future behaviour of the monitored state machines.

\item \textbf{Appliance Naming:} In the final step, the algorithm attempts to
  assign each observed FSM to an actual appliance in the system. For this, Hart
  recommends Bayesian, maximum-likelihood-multiple-hypothesis or other methods
  from detection theory.
\end{enumerate}

\section{Evaluation Metrics for NILM Algorithms} \label{sec:metrics}

To evaluate the performance and quality of a load-disaggregation algorithm,
numerous accuracy metrics are reported in NILM literature. We have identified no
common agreement in the community on which metrics are superior or more suitable
to certain approaches. This in particular can bar the way to a quantitative
comparison of different NILM techniques. We will nevertheless outline the
general categories of evaluation metrics and briefly touch upon specific
examples. We also wish to note and laud up-front the work of Batra, Kelly
\emph{et al.} in publishing NILMTK, an open-source toolkit for evaluating NILM
algorithms that implements many of the metrics discussed below \cite{nilmtk}
\footnote{http://nilmtk.github.io}.

\subsection{Event-Based Metrics}

Event-based metrics focus on evaluating the detection of state-changes in the
operation of appliances. More specifically, it is analysed how well an algorithm
can identify an appliance switching on or off \cite{Aiad201696}. Event-based
metrics make use of the terms \emph{True Positive} (TP), \emph{True Negative}
(TN), \emph{False Positive} (FP) and \emph{False Negative} (FN). TP refers to
the number of times a device is correctly identified as \emph{on}, while TN is
the count of correctly captured \emph{off} events. Conversely, a FP event stands
for the case when an active state was reported, when the device was in fact not
consuming power. FN is defined analogously. Given these quantities, the
following evaluation metrics can be calculated:

\begin{itemize}
\item \textbf{Precision}: The ratio of samples an appliance was correctly
  detected as active:
  \begin{equation}
    Precision = \frac{TP}{TP + FP} \in [0,1]
  \end{equation}

\item \textbf{Recall}: The ratio of samples the algorithm captured an actual
  turn-on event:
  \begin{equation}
    Recall = \frac{TP}{TP + FN} \in [0,1]
  \end{equation}

\item \textbf{Accuracy}: The ratio of samples the algorithm correctly identified
  the actual state of an appliance:
  \begin{equation}
    Accuracy = \frac{TP + TN}{TP + TN + FP + FN} \in [0,1]
  \end{equation}
\end{itemize}

\subsection{Non-Event-Based Metrics}

Non-event-based metrics analyse how well a load-disaggregation algorithm is able
to compute and assign the power-consumption of individual appliances
\cite{nilmtk}. One drawback of such metrics is that may produce favourable
quantitative results for appliances that are inactive for long spans of time,
shadowing the ability of an algorithm to identify a device's power draw in one of
its active operating states. Given that most household appliances such as
televisions or microwaves are inactive for the majority of the day, an algorithm
may score very highly on such a non-event-based metric simply by assigning zero
power consumption to all appliances. As stated in \cite{kolter2011}, this class
of metrics is nevertheless applicable to a wider range of algorithms, as it does
not mandate the extraction of edges from the aggregate power signal.

One such evaluation criterion, as reported in \cite{nilmtk}, is the root mean
square error (RMSE). More specifically, the RMSE between the identified power draw
and the actual power draw of an appliance is given by the equation:
\begin{equation}
  RMSE = \sqrt{\frac{\sum_t(y_t^{(n)} - \hat{y}_t^{(n)})^2}{T}}
\end{equation}
Where $T$ denotes the number of samples recorded, $y_t^{(n)}$ the
assigned power draw of the $n$-th appliance at time instant $t$, and $\hat{y}$
the actual power draw of that appliance at that time instant. Another variation
of this metric, as employed by \cite{Egarter2015}, is to normalise the RMSE over
the range of all recorded samples, yielding the normalised RMSE (NRMSE):
\begin{equation}
  NRMSE = \frac{RMSE}{max(y_t^{(n)}) - max(y_t^{(n)})}
\end{equation}

\subsection{Overall Metrics}

Lastly, another class of metrics exists which evaluates the overall performance
of a load-disaggregation algorithm. In the simplest scenario, one would rate the
quality of a NILM technique by calculating the percentage of the total power
assigned to a specific appliance for the entire duration of the experiment,
relative to the aggregate consumption of all appliances \cite{kolter2011}. One
would then inspect the difference between this value and the ground truth
quantity, calculated identically. Let $Y_t = \sum_n y_t^{(n)}$ denote the total
aggregate power consumption \emph{estimated} at time $t$, and $\hat{Y}_t$ the
ground-truth equivalent. The average deviation in the ratio between assigned and
actual power consumption may be calculated as:

\begin{equation}
  \frac{1}{N}\sum_n(\left|\frac{\sum_t(y_t^{(n)})}{\sum_t(Y_t)} - \frac{\sum_t(\hat{y}_t^{(n)})}{\sum_t(\hat{Y}_t)}\right|)
\end{equation}

\section{Datasets for NILM Research} \label{sec:datasets}

There exist a number of publicly available datasets to test and train the
various aforementioned learning approaches. We have picked two such datasets for
the purpose of further discussion and analysis of their characteristics.

\subsection{The Reference Energy Disaggregation Data Set}

One of the most detailed and widely used datasets is the Reference Energy
Disaggregation Data Set (REDD) \cite{kolter2011}. It was released to the public
for purposes of load-disaggregation research by MIT in 2011. REDD is composed of
aggregated as well as sub-metered, device-specific power-consumption data gathered from 10 homes monitored over a period of two weeks. In total, it
contains a combined 119 days of measurements spread over 1 terabyte of
data. High as well as low frequency samples were extracted from three separate
sources:

\begin{enumerate}
\item Mains electricity, sampled at $\SI{15}{kHz}$.
\item Labeled circuits, recorded at $\SI{0.5}{Hz}$.
\item Plug-level monitors, gathered at $\SI{1}{Hz}$.
\end{enumerate}

While REDD finds its use in sources such as \cite{Egarter2015} and \cite{eps364263}, it is mentioned in \cite{Aiad201696} that the total aggregate
of individual appliances' power consumptions at certain time points differs from
the mains measurement reported in the dataset. As also noted in
\cite{Aiad201696}, this inconsistency indicates the existence of additional,
unmeasured devices.

\subsection{GREEND}

Another publicly available dataset is GREEND, released by Monacchi \emph{et al}. in
2014 \cite{monacchi2014}. GREEND consists of active power measurements of
nine houses for a continuous duration of one year, collected in the region of
Carinthia, Austria and Friuli-Venezia, Italy. Each house contains an average of
nine appliances, totaling 79 different power-consumption sources. It claims to
be the first power-consumption dataset for Austria and Italy measured at a 1 Hz
sampling-frequency. The long period of measurement in particular makes GREEND
very attractive for NILM research, allowing investigation and modeling of
seasonal consumption behaviour. Other datasets such as REDD, lasting
approximately two weeks \cite{kolter2011}, or BLUED \cite{englert:2013},
spanning 8 days, comprise much less data, in comparison.

\section{Recent approaches} \label{sec:newideas}

A more recent approach suggests a rethinking of NILM itself and introduces a new
way of implementing the algorithm. The authors of \cite{barker2014nilm}
introduce a new modelling of NILM in a application-centric way. The approach
demands for real-time processing right after metering, which is termed
\emph{online} NILM. Basically it is suggested to divide NILM into three steps:
device detection, modelling, and device-tracking. The novelty of this approach
is that device detection and modelling are usually said to be offline tasks as
part of algorithm-training. For the method proposed, device recognition and
monitoring is implemented online and in real-time. Smart meters would thus
transmit the measurement data immediately to the cloud or server. For example,
the online service could be hosted by the power utility itself, improving its
immediate ability to forecast future power consumption. The crux of this idea
lies in computation. Such a system would have to perform NILM across hundreds of
households in real-time. We identify this in particular as a challenge.

The most common learning algorithms used for load-disaggregation today rely on
optimisation or Factorial Hidden Markov Models (FHMM). In \cite{kelly2015},
Kelly et al. very recently employed a novel learning approach based on
artificial neural networks (ANN). For this, the authors implemented three
separate ANN architectures. The first is a \emph{recurrent neural network},
which learns appliance features on a training dataset to then estimate the
appliance consumption level given an aggregate sample. The second architecture
utilises a \emph{denoising autoencoder} (dAE), often used for signal
reconstruction and denoising, such as for removing grain from an old image or
reverberation from an audio track. For the dAE, the learning task is to extract
a device's load from an aggregate sample, by viewing the consumption of other
appliances as the signal's unwanted noise component. Lastly, a standard neural
network was used to regress start and end time as well as power consumption for
each activation of a device. We note that little research has been done on the
application of these modern machine learning techniques to NILM. Yet,
\cite{kelly2015} shows that neural networks beat conventional
load-disaggregation algorithms in almost every metric, inviting further
investigation into these promising new learning approaches.

\section{Conclusion} \label{sec:conclusion}

In this paper we discussed the concept of NILM, appliance models, appliance
signatures and learning methods as well as recent improvements and trends in the
field of load-disaggregation. We are certain further research is necessary. One
fundamental question posed is \emph{where} and on what platforms data processing
and NILM algorithms are performed. The first and simplest option is the
measurement device itself, meaning the smart meter or metering units installed
in the household. This would require sophisticated hardware that is capable of
performing NILM in real-time. In general, such hardware is more expensive and
energy-consuming than conventional measurement devices. Therefore another
approach is to perform data processing on a device in the home
network. Single-board computing devices such the Raspberry
Pi\footnote{https://www.raspberrypi.org} or
BeagleBoard\footnote{https://beagleboard.org} could be well suited to such a
task. With the surge of cloud-computing services in recent years, the employment
of such an online service becomes a possibility as well. However, along with
data-transmission across networks, away from the home and into the cloud,
security concerns will and must be raised. We estimate that the majority of the
population would feel unease in sending their private household data to external
servers.\\In conclusion, we would like to re-emphasise our belief in the very
certain potential of load-disaggregation techniques to improve the consumption
patterns of individuals and reduce energy wastage in the grid. At the same time,
we acknowledge that non-intrusive load-monitoring is still a very open and
ongoing field of research and that no current approach is perfect. We express
our hope that this will change in the near future.

\bibliographystyle{IEEEtran}
\bibliography{bibfile}

\end{document}